\documentstyle[12pt]{article}

\begin{document}

\baselineskip=6.8mm

\newcommand{\TeV}{\,{\rm TeV}}
\newcommand{\GeV}{\,{\rm GeV}}
\newcommand{\MeV}{\,{\rm MeV}}
\newcommand{\keV}{\,{\rm keV}}
\newcommand{\eV}{\,{\rm eV}}
\newcommand{\Tr}{{\rm Tr}\!}
\renewcommand{\arraystretch}{1.2}
\newcommand{\be}{\begin{equation}}
\newcommand{\ee}{\end{equation}}
\newcommand{\bea}{\begin{eqnarray}}
\newcommand{\eea}{\end{eqnarray}}
\newcommand{\ba}{\begin{array}}
\newcommand{\ea}{\end{array}}
\newcommand{\bmat}{\left(\ba}
\newcommand{\emat}{\ea\right)}
\newcommand{\refs}[1]{(\ref{#1})}
\newcommand{\ler}{\stackrel{\scriptstyle <}{\scriptstyle\sim}}
\newcommand{\ger}{\stackrel{\scriptstyle >}{\scriptstyle\sim}}
\newcommand{\lag}{\langle}
\newcommand{\rag}{\rangle}
\newcommand{\ns}{\normalsize}
\newcommand{\cm}{{\cal M}}
\newcommand{\gr}{m_{3/2}}
\newcommand{\p}{\partial}
\def\tl{{\tilde{l}}}
\def\tL{{\tilde{L}}}
\def\bd{{\overline{d}}}
\def\tL{{\tilde{L}}}
\def\a{\alpha}
\def\b{\beta}
\def\g{\gamma}
\def\c{\chi}
\def\d{\delta}
\def\D{\Delta}
\def\db{{\overline{\delta}}}
\def\Db{{\overline{\Delta}}}
\def\e{\epsilon}
\def\l{\lambda}
\def\n{\nu}
\def\m{\mu}
\def\nt{{\tilde{\nu}}}
\def\p{\phi}
\def\P{\Phi}
\def\x{\xi}
\def\r{\rho}
\def\s{\sigma}
\def\t{\tau}
\def\th{\theta}
\renewcommand{\Huge}{\Large}
\renewcommand{\LARGE}{\Large}
\renewcommand{\Large}{\large}

\begin{titlepage}
\title{\bf Low $\a_{strong}(M_Z)$, Intermediate Scale SUSY
SO(10) and Its Implications\\
                          \vspace{-4cm}
                          \hfill{\ns \\}
                          \hfill{\ns UMD-PP-95-138\\}
                          \hfill{\ns May 1995}
                          \vspace{3cm} }

\author{ Biswajoy Brahmachari$^\dagger$
         and Rabindra N. Mohapatra$^*$\\
  {\ns\it $^\dagger$International Center for Theoretical Physics}\\
  {\ns\it P.~O.~Box 586, 34100 Trieste, Italy.} \\
  {\ns\it $^*$ Department of Physics, University of Maryland}\\
  {\ns\it College Park, Maryland, 20742.} }

\date{}
\maketitle
\vspace{2cm}
\begin{abstract} \baselineskip=7mm
{\ns }  We show that one of the ways of obtaining consistency between the
idea of supersymmetric grand unification and an apparent low value of
$\alpha_{strong}(M_Z)\simeq .11$ indicated by several low energy experiments
is to have an intermediate scale corresponding to a local $B-L$ symmetry
breaking around the mass scale of $10^{10}$ to
 $10^{12}$ GeV. We discuss the realization
of this idea within the framework of supersymmetric $SO(10)$ grand unified
theories with spectra of particles derivable from simple superstring-inspired
versions of this model. We then study the $b-\tau$
mass unification within this class of models and show that due to the
influence of new gauge and Yukawa interactions beyond the intermediate
scale, the prediction of the $b$-quark mass comes out well within the
presently accepted values. We also discuss an un-orthodox class of SUSY
models inspired by some theoretical considerations having
two pairs of Higgs doublets at low energy and show that they also
can lead to unification with intermediate scales and low $\alpha_{s}(M_Z)$
as desired.

\end{abstract}

\thispagestyle{empty}
\end{titlepage}

The high precision measurements of the gauge couplings $\alpha_1,\alpha_2$ and
$\alpha_{strong}$ at LEP combined with the coupling constant evolution
dictated by the minimal supersymmetric standard model, have
 in the past three years led to speculations that the present
data while providing overwhelming support for the standard model may
in fact be indicating that the next level of physics consists of a single
scale supersymmetric grand unified theory, with new physics beyond
supersymmetry appearing only at the scale of $10^{16}$ GeV\cite{amaldi}. This
has generated a great deal of excitement and activity in the area
of supersymmetric grand unified theories(SUSY GUT).
There are however reasons to believe that there could
be  new physics starting below $10^{16}$ GeV,
perhaps somewhere around $10^{10}$ to $10^{12}$ GeV or so
based on attempts to understand the hot dark
matter of the universe and to solve the strong CP problem etc.
Since the idea of SUSY GUTs
is so attractive, it is then important to ask whether an
intermediate scale in the $10^{12}$ GeV range fits within
the framework of SUSY GUTs.

 In this letter, we show that if
the value of the QCD fine structure constant $\alpha_{s}(M_Z)$
at the weak scale
turns out to be around $0.11$ as is indicated by several low
energy experiments, then a simple and interesting
 way to reconcile it with the idea
of SUSY GUTs is to have an intermediate scale around $10^{10}$ to
$10^{12}$ GeV.
Thus it may be argued that a lower value of $\alpha_{s}(M_Z)$ may indeed
provide an automatic reconciliation between the idea of SUSY GUTs
and the idea of an intermediate scale motivated
by reasons completely unrelated to it. There are of course scenarios
where a low $\a_{s}(M_Z)$ is consistent with intermediate scales
other than the ones given above; we do not consider them here since
they may be less well motivated from other considerations.

Before beginning the discussion of the intermediate scale GUT models,
let us briefly review the situation with respect to the value of
$\alpha_{s}(M_Z)$. A very useful summary of the issues have been given
in a recent paper by Shifman\cite{shifman}, who has suggested
that the discrepancy
between the higher values of $\approx 0.125$ for
$\alpha_{s}(M_Z)$ derived from LEP data on the
one hand and the
low energy data such as deep inelastic electron scattering,
lattice calculations involving the upsilon and the $J/\Psi$ system
etc on the other should be considered to
be an indication of the presence
of new physics\cite{langa}.
If this new physics is identified with the supersymmetric
version of the standard model(MSSM), then one can attempt to do a global
fit to all LEP data and see if indeed the high value of $\alpha_{s}(M_Z)$
indicated there is lowered. Such an analysis has been carried
out recently by Kane, Stuart and Wells\cite{kane}, who show that
if the stop and the chargino masses in the MSSM are kept below a 100 GeV,
then there are new contributions to the $Z\rightarrow b\overline{b}$
decay which increase its decay width.
In the presence of these contributions,
the global fit to LEP data indeed leads to a value for $\alpha_{s}(M_Z)
\simeq .112$ which is what the low energy data give for this parameter.
It should be noted that there are a number of other suggestions that
could also lead to a higher value for the $Z\rightarrow b\overline{b}$ width.
It could very well be that one of these scenarios rather than the SUSY
contribution is at the real heart of the problem. But for our discussion of
unification, it important that either the experimental value of the
$Z\rightarrow b\overline{b}$ come down or that the supersymmetric scenario
provide an explanation for its enhancement over the standard model value.

The question then is what this implies for the unification of coupling
constants. It is by now well-known that in the single scale unification
theory, one predicts a value of $\alpha_{s}(M_Z)\simeq .125$ or so. In a recent
letter, Bagger et al\cite{bagger} have shown that in the minimal SU(5) type
theories, the inclusion of threshold corrections do not change this situation.
Therefore one has to go beyond the simple single scale canonical
GUT models to accommodate the lower value for $\alpha_{s}(M_Z)$.

One possibility is to espouse an alternative string type unification rather
than the conventional SUSY GUT theory
as has been done by Shifman and Roszkowski
\cite{shifman2} and give up the unification of gaugino masses.

Our goal in this letter is to explore alternatives within the framework of
the GUT hypothesis which can achieve the same goal i.e.
accommodate a lower $\alpha_{s}(M_Z)$ of about $0.11$ while leading to
eventual gauge coupling unification.
We show that if a SUSY GUT theory based on SO(10) is allowed to have an
intermediate scale in the interesting range of $10^{10}$ to
$10^{12}$ GeV or so,
one can construct scenarios which lead to $\alpha_{s}(M_Z)\simeq 0.11$
keeping all other phenomenological implications consistent while
adding the extra desirable features mentioned earlier.

In a general analysis of the SUSY SO(10) models with intermediate
scales, it was noted recently
that\cite{leemoh} that there are several instances where
 the intermediate scale is consistent
with grand unification only if the value of $\alpha_{s}(M_Z)\simeq .11$ or so.
We pursue this alternative here and study the possibility of $b-\tau$
unification in this class of models.  In addition to confirming the results
of \cite{leemoh}, we obtain the following new results:
(i) we show that $b-\tau$ unification also prefers a lower
value of $\alpha_{s}$; (ii) we then
 present a new example of an unorthodox
SUSY model at low energy which can lead to a grand unified SO(10) model with
intermediate scale. The new model we have in mind is the one with two pairs
of $(H_u,H_d)$ in the low energy supersymmetric version of the standard model
as against only one pair in the conventional MSSM.
These models have potentially richer phenomenology
then the conventional MSSM
and may have certain advantages in incorporating spontaneous CP violation
into the supersymmetric standard model.

 We will work within the class of SO(10) models
suggested by the superstring theories with level two
 compactification\cite{lykken}, so that all particles assumed in our spectra
arise from {\bf 10}, ${\bf 16}+\overline{\bf 16}$ ,and {\bf 45} dimensional
representations. The generic kind of theories we will be interested in
will consist of two {\bf 10} dimensional, one or two {\bf 45} dimensional
multiplets
and several pairs of ${\bf 16}+\overline{\bf 16}$ at the GUT scale $M_U$.
Only a small subset of these multiplets will be assumed to remain light
below $M_U$ down to the intermediate scale $M_{B-L}$ and an even smaller
subset below $M_{B-L}$ down to $M_W$. We will define two classes of
models depending on the particle spectra below $M_{B-L}$. Both the classes
will of course contain the usual matter multiplets which will give the
quarks and leptons of the three generations but will be distinguished
by whether they have one or two pairs of Higgs doublet superfields:
$(H_u, H_d)$. The model with one pair is of course the conventional MSSM
and will be denoted as such; the model with two pairs will be called
MSSM2. It is of course well known that MSSM2 is inconsistent with the
idea of single scale grand unification but we will show that once we allow
for the existence of an intermediate $B-L$ symmetry scale, it is in fact
consistent with a lower $\alpha_{s}(M_Z)$ which is the main focus of this
paper. Let us consider these cases separately.

{\underline {\it I. The MSSM case:}}

It was pointed out in Ref.\cite{leemoh} that under the above conditions, the
only particle spectrum between $M_U$ and $M_{B-L}$ which can lead to
$\alpha_{s}(M_Z)\simeq 0.11$ is the one  with two pairs of (2,2,0,1) and
five pairs of $(1,2,1,1)+(1,2,-1,1)$ multiplets under the group
$SU(2)_L\times SU(2)_R\times U(1)_{B-L}\times SU(3)_c$. Our calculations have
confirmed this result and we find in the one loop approximation that
$\alpha_{s}(M_Z)=0.11$, $M_U\simeq 2.05\times 10^{15}$ GeV and $M_{B-L}\simeq
3.47\times 10^{12}$ GeV. As shown in Ref.\cite{leemoh}, inclusion of the
two loop corrections change this result only very slightly.
 The graph displaying the evolution of the three
gauge couplings in the one loop approximation is given in Fig.1.
 The lower value of the $M_U$ leads to a gauge
mediated proton lifetime ( for the decay mode $p\rightarrow e^+\pi^0$ )
of $.62\times 10^{33\pm.7}$ years up to threshold corrections. It is
interesting that this is within the reach of the Super Kamiokande experiment.
The Higgsino mediated proton decay amplitude in this case must be
suppressed by some additional mechanism\cite{babubarr}

  Let us now address the question of predicting
the mass of the  top and the bottom quark masses
in these theories: the first by using the idea of infra-red fixed
point of the top Yukawa coupling assumed to be large at the GUT
scale \cite{hill} and the second by using the property of these
models that at the GUT scale, one has $m_b(M_U)=m_{\tau}(M_U)$.
In this class of models where the $B-L$ symmetry is broken at the
intermediate scale, the conventional approach to both these questions
will be modified\cite{brahmachari,ross,leemoh} due to new evolution
equations for the Yukawa couplings above $M_{B-L}$.

It is widely known
that in SUSY GUTs with one step breaking predict a large value of $m_b$
for the major part of the parameter space \cite{barg}. The study of
b-$\tau$ unification including a right handed neutrino has also been
performed \cite{fv}. However, in this study no new
gauge interactions beyond the intermediate scale has was considered
and due to renormalization effects of the new Yukawa coupling,
a 10-15 \% increase in
the mass of the b-quark was obtained. We will show that after the
inclusion of the new left-right symmetric gauge and Higgs interactions
at $M_{B-L}$
surviving from an SO(10) GUT, the running of the b-quark Yukawa coupling
is altered and as a result an attractive reconciliation with the
experimental measurements can be achieved. To discuss these,
let us recall that since there are two Higgs bidoublets (denoted by
$\phi_1$ and $\phi_2$) above $M_{B-L}$,
the Yukawa superpotential of the quarks and leptons are given by:
\be \label{a1}
{\cal W_Y} = h_{Q_1} Q^T\tau_2\phi_1 Q^c + h_{Q_2} Q^T\tau_2\phi_2 Q^c
+ h_{L_1} L^T\tau_2 \phi_1 L^c + h_{L_2} L^T\tau_2\phi_2 L^c + h.c.\;
\ee
(where we have denoted the quarks and leptons by the obvious notation
$Q, Q^c$ and $L,L^c$). At the GUT scale, we have $h_{Q_1}(M_U)=h_{L_1}(M_U)$
and $h_{Q_2}(M_U)=h_{L_2}(M_U)$. It is possible
 to construct doublet triplet splitting mechanisms \cite{babu}
where the $H_u$ and $H_d$ arise from $\phi_1$
and $\phi_2$ separately. In this case, the model has b-$\tau$ unification
at the GUT scale and we can study whether the bottom mass is predicted
correctly at low energies. The  evolution
equations above $M_{B-L}$  in this case are given by \cite{leemoh}. We are
rewriting them in the notation of \cite{jones}, as,
\be \label{a2}
{{d Y_a}\over{dt}} = Y_a~[S_{ab} Y_b -Z_{ai}\alpha_i]. \;
\ee
The variable t is given by,
\be
t={1 \over 2 \pi}~ln{ \mu \over GeV},
\ee
where $\mu$ denotes the renormalization scale, and,
$S$ is a $ 4 \times 4$ matrix and $Z$ is a
$4\times 4$ matrix given by:
\be \label{a3}
S = \bmat{cccc} 7 & 4 & 1 & 0\\ 4 & 7 & 0 & 1\\ 3 & 0 & 5 & 4\\
0 & 3 & 4 & 5 \emat ,\;
\ee
and
\be \label{a4}
Z = \bmat{cccc} 1/6 & 3 & 3 & 16/3\\ 1/6 & 3 & 3 & 16/3 \\ 3/2 & 3 & 3 &
0\\ 3/2 & 3 & 3 & 0\emat .\;
\ee
In the above equations, we have chosen the basis $ ( Y_{Q_1}, Y_{Q_2},
Y_{L_1}, Y_{L_2} )$  with  $Y_i = h^2_i/{4\pi}$ and the basis for the
fine structure constants is the one corresponding to $( U(1)_{B-L},
SU(2)_L, SU(2)_R, SU(3)_c )$. We point out that below
the scale $M_{B-L}$ the right-handed neutrino contributions must be absent
since this particle decouples from low energy physics.
The equations for the Yukawa couplings also
change and acquire the standard form \cite{hill}. We
have numerically solved these equations for a large range
 of values for $Y_{Q_2}=Y_{L_2}=Y_2$. We have kept the top Yukawa
coupling $Y_{Q_1}=Y_{L_1}=Y_1$ at $M_U$
to be equal to 1 and obtained in different cases the values
for the ratio $R_{b/\tau}(M_t)\equiv
m_b(M_t)/m_{\tau}(M_t)$ which are summarized in Table. 1. We
point out
that for $\alpha_{s}(M_Z)\simeq 0.11$, the low energy values of $m_b$ and
$m_{\tau}$ when extrapolated to the scale $M_Z$ gives an $R_{b/\tau}(M_t)
\simeq 1.7~to~1.9$ for values of $m_b(m_b)= 4.1$ to $4.5$ GeV. Thus the
values of $m_b$ obtained by us in these models are quite consistent.
The corresponding predictions for the top mass $m_t(pole)=m_t(M_t)~[1+{4
\alpha_s \over 3 \pi}]$ lies between 177 to 199 GeV which again is also
consistent with data. The value of $tan\beta$ is determined from the
measured value of the tau lepton mass.
In Fig.2, we display the evolution of the Yukawa couplings for the case
of $Y_{Q_2}(M_U)= Y_{L_2}(M_U)=0.1$ as typical example.

\begin{table}[htb]
\begin{center}
\[
\begin{array}{|c||c||c||c||c||c||c||c||c|}
\hline
Y_1&Y_2&h_t(M_t)&h_b(M_t)&h_\tau(M_t)&
tan\beta&m_b(M_t)&m_t(M_t)&{m_b(M_t) \over m_\tau(M_t)}\\
\hline
1&1         &1.010&0.96&0.62&60.43&2.77&176.83& 1.56 \\
1&10^{-1}   &1.060&0.84&0.52&51.26&2.85&184.46&1.60  \\
1&10^{-2}   &1.094&0.460&0.270&26.7&3.01&190.34&1.69  \\
1&10^{-3}   &1.103&0.160&0.095&9.25&3.06&190.90&1.72  \\
1&10^{-4}   &1.104&0.054&0.030&2.80&3.07&181.03&1.73  \\
1&{1 \over 2}10^{-4}   &1.104&0.037&0.021&1.85&3.08&169.16&1.73 \\
\hline
\end{array}
\]
\end{center}
\caption{The values of $h_t(M_t)$, $h_b(M_t)$, $h_\tau(M_t)$ and
calculated by RGE for
$\alpha_s=0.11$ in the MSSM case. The prediction of the
masses $m_b$ and $m_t$ at the scale
$M_t$ has been quoted in GeV. $M_t$ is defined as 170 GeV. $tan \beta$
has been calculated assuming $m_\tau(M_Z)=1.777$ GeV.} \label{table1}
\end{table}

\newpage
\noindent{\underline {\it II. The MSSM2 Model:}}

In this section, we present the new possibility for the low energy
supersymmetric model which also leads to a lower $\alpha_{s}(M_Z)$.
For this model , the particle spectra below $M_{B-L}$ consists of
an extra pair of Higgs doublet superfields $(H_u,H_d)$ in addition to
the particles already present in the MSSM. We call this model MSSM2
from now on. Above the scale $M_{B-L}$, we assume the additional
particles to be: one (1,1,0,8) multiplet, four pairs of $(1,2,1,1)
+(1,2,-1,1)$ , two bidoublets (2,2,0,1) in the basis $SU(2)_L \times
SU(2)_R \times U(1)_{B-L} \times SU(3)_c $.
Below the $M_{B-L}$ scale, we assume this theory to consist of two pairs
of the Higgs doublets of MSSM2. Note that the (1,1,0,8) multiplet
could arise from the SO(10) multiplet {\bf 45}.
We then find that there is coupling constant
unification with $\alpha_{s}(M_Z)=.11$, $M_{B-L}\simeq 1.52\times 10^{11}$ GeV
and $M_U\simeq 1.72\times 10^{16}$ GeV. All these numbers are evaluated
in the one loop approximation; we expect them to change by 1 or $2\%$
due to two loop corrections. The one loop running for this case is shown in
Fig.3. The gauge mediated proton decay in this case is suppressed as in the
usual single scale GUT models.

Before exploring the implications of $b-\tau$ Yukawa unification
in this case, we assume that due to some discrete symmetry,
below $M_{B-L}$, only one
pair of the left-handed doublets remains coupled to the fermions and another
pair just "floating around" to help with soft CP violations etc.
The rest of the discussion
similar to the one  in the previous section except that
here we have a different evolution of the gauge couplings above the $M_{B-L}$
scale. We have also numerically analyzed the Yukawa evolution equations
in this case; we give it in Fig.4 and again, we find that the bottom quark
mass comes out in very good agreement with experiment and the predicted
value of $m_t(M_t)$ is given in Table 2. $m_t(pole)$ is in the range 173
to 198 GeV in this case.

\begin{table}[htb]
\begin{center}
\[
\begin{array}{|c||c||c||c||c||c||c||c||c|}
\hline
Y_1&Y_2&h_t(M_t)&h_b(M_t)&h_\tau(M_t)&
tan\beta&m_b(M_t)&m_t(M_t)&{m_b(M_t) \over m_\tau(M_t)}\\
\hline
1&1         &0.99&0.960&0.59 &57.81   &2.87&173.90&1.61 \\
1&10^{-1}   &1.05&0.830&0.50 &48.78   &2.97&182.65&1.67  \\
1&10^{-2}   &1.09&0.460&0.26 &25.41   &3.17&189.11&1.78  \\
1&10^{-3}   &1.10&0.160&0.09 &8.81    &3.23&189.69&1.82  \\
1&10^{-4}   &1.10&0.052&0.023 &2.65   &3.24&178.80&1.82  \\
1&{1 \over 2}10^{-4}   &1.104&0.037&0.020&1.74&3.24&165.70&1.82 \\
\hline
\end{array}
\]
\end{center}
\caption{The values of $h_t(M_t)$, $h_b(M_t)$, $h_\tau(M_t)$ and
calculated by RGE for $\alpha_s=0.11$ in MSSM2 case. The
prediction of the masses $m_b$ and $m_t$
at the scale $M_t$ has been quoted in GeV. $M_t$ is
defined as 170 GeV. $tan \beta$ has been calculated assuming
$m_\tau(M_Z)=1.777$ GeV.} \label{table3}
\end{table}

 These models may also have
advantages for introduction of spontaneous CP violation\cite{garisto}
 and have a potentially rich Higgs phenomenology\cite{haber}. We hope to
explore these questions.

We now briefly discuss a model, which we call MSSM3, that reduces to MSSM2
below $M_{B-L}$. The model we present arises by
 adding an extra pair of (2,1,1,1)+(2,1,-1,1) multiplets,
above the scale $M_{B-L}$.
 Denoting them by $\chi_L+\overline{\chi}_L$, and the corresponding
righthanded doublets by $\chi_R+\overline{\chi}_R$, we can have couplings
of the form $\overline{\chi}_L \overline{\chi}_R \phi$
 ( where $\phi\equiv (2,2,0,1)$). Let us choose the relevant $\phi$ to
be the $\phi_1$ of Eq.1. Once
$\overline{\chi}_R$ acquires a vev the down type doublet in
 the $\phi$ pairs up with $\overline{\chi}_L$
to acquire mass of order $M_{B-L}$ and the accompanying ${\chi}_L$
which is uncoupled to quarks  and leptons remains as an $H^{\prime}_d$
at low energies. Now note that in Eq.1, the $\phi_2$ couplings lead
to bottom quark mass and we can choose this coupling to be small
( for small $tan\beta$). Even though the $\phi_2$ contains the
second $H_u$ of MSSM2, because of its small Yukawa coupling,
it leaves the Yukawa evolution discussion of the previous
section remains unchanged (except for the effect of changes in the evolution
of the gauge couplings). We have studied the Yukawa and gauge coupling
evolution in this model and they are given in fig. 5 and 6
and some representative numbers in the table 3.
 In this model, the value of $M_U
\simeq 3.55\times 10^{15}$ GeV and the value of $M_R\simeq 4.81\times 10^9$
GeV. The Yukawa unification also works in this case.
Further details of the model will be the subject of a longer paper under
preparation.

\begin{table}[htb]
\begin{center}
\[
\begin{array}{|c||c||c||c||c||c||c||c||c|}
\hline
Y_1&Y_2&h_t(M_t)&h_b(M_t)&h_\tau(M_t)&
tan\beta&m_b(M_t)&m_t(M_t)&{m_b(M_t) \over m_\tau(M_t)}\\
\hline
1&10^{-4}   &1.104&0.050&0.027&2.51&3.23&178.23&1.81  \\
1&{1 \over 2}10^{-4}   &1.104&0.035&0.019&1.63&3.21&163.57&1.81 \\
\hline
\end{array}
\]
\end{center}
\caption{The values of $h_t(M_t)$, $h_b(M_t)$, $h_\tau(M_t)$ and
calculated by RGE for
$\alpha_s=0.11$ in the MSSM3 case. The prediction of the
masses $m_b$ and $m_t$ at the scale
$M_t$ has been quoted in GeV. $M_t$ is defined as 170 GeV. $tan \beta$
has been calculated assuming $m_\tau(M_Z)=1.777$ GeV.} \label{table3}
\end{table}

\vspace{4mm}

In conclusion, we have shown that a simple way to reconcile a lower
value of $\alpha_{s}(M_Z)$ with the SUSY GUT hypothesis is to admit the
existence of an intermediate scale corresponding to local $B-L$ symmetry.
While the value of the intermediate scale depends on the choice of
light Higgs multiplets, we have focussed on scenarios where this
scale is around $10^{10}$ to $10^{12}$ GeV or so since they have other
independent physical motivations. We have outlined three classes of
scenarios which achieve this goal; one with the usual MSSM structure at
low energies and a second one which has two pairs of Higgs doublets in
the low energy domain. We have also shown that such gauge and
Higgs structures above the intermediate scale predicted by the unification
of gauge couplings influences the running of the b quark Yukawa
coupling in the proper direction. As a result, a correct prediction of the
$m_b(m_b)$ can be obtained which is otherwise difficult to achieve in MSSM
with single scale GUT.
Furthermore, our scenario also predicts top quark mass in the
appropriate
range when the top quark Yukawa coupling is in its fixed point domain
at the unification scale. We have also outlined a scenario where one
can realize the scnario of MSSM2 in a gauge model.

{\bf Acknowledgement:}

The work of R.N.M. was supported by the National Science foundation grant
no. PHY9421385. He thanks A. Rasin and M. Masip for discussions of
the MSSM2 model. B. B. thanks the Physics Department of the University of
Maryland for hospitality and financial support during a short visit.

\section*{Figure Captions}

\noindent Figure 1. The evolutions of various $\alpha^{-1}_i$ and the
unification of gauge couplings in model 1. We have chosen ${{1}\over{2\pi}}
ln{{\mu}\over{GeV}}$ as the variable represented along the x-axis in
this and all subsequent figures.\\

\noindent Figure 2. The evolution of Yukawa couplings in model 1. The
different lines from top to bottom represent the evolution of $Y_t,
Y_{\nu}, Y_b $ and $Y_{\tau}$ respectively. \\

\noindent Figure 3. The evolution of various $\alpha^{-1}_i$ and the
unification of gauge couplings in model 2.\\

\noindent Figure 4. The evolution of Yukawa couplings in model 2 with
the lines representing the cases as in Fig.2 .\\

\noindent Figure 5. The evolution of th various $\a^{-1}_{i}$ and the
unification of gauge couplings for MSSM3.\\

\noindent Figure 6. The evolution of Yukawa couplings in a typical case
for MSSM3.\\

\end{document}